\documentclass{article}
\usepackage{graphicx}

\begin{document}

\begin{center}
 {\bf The companion properties of SNe Ia from the single degenerate
model}

\author[{MENG XiangCun$^{\rm 1}$ \& YANG WuMing$^{\rm 1,2}$}

{$^{\rm1}$}{School of Physics and Chemistry, Henan Polytechnic
University, Jiaozuo, 454000, China;}

{$^{\rm2}$}{Department of Astronomy, Beijing Normal University,
Beijing 100875, China.}
\end{center}

\begin{center}

{\begin{abstract} Although Type Ia supernovae (SNe Ia) play very
important roles in many astrophysical fields, the exact nature of
the progenitors of SNe Ia is still unclear. At present, the single
degenerate (SD) model is a very likely one. Following the
comprehensive SD model developed by Meng \& Yang (2010), we show
the properties of SNe Ia companions at the moment of the supernova
explosion. The results may provide help in searching for companion
stars in supernova remnants. We compared our results with the
companion candidate Tycho G of Tycho's supernova and found that
integral properties of the star (mass, space velocity, radius,
luminosity and effective temperature) are all consistent with
those predicted from our SD model with the exception of the
rotational velocity. If Tycho G was the companion star of Tycho's
supernova, an interaction between supernova ejecta and the
rotational companion might be a key factor to solve the
confliction, and then it could be encouraged to do a detailed
numerical simulation about the interaction.
\end{abstract}}
\end{center}\vspace*{-0.6cm}
\begin{center}
\end{center}

\section{Introduction}\label{sect:1}
As wonderful distance indicators, Type Ia supernovae (SNe Ia) play
an important role in cosmology and have been used successfully to
determine cosmological parameters, resulting in the discovery of
the accelerating expansion of the Universe [1,2]. However, the
exact nature of SNe Ia progenitors is still unclear, especially
the progenitor system [3,4]. Regarding the nature of the mass
accreting WD companions, two competing scenarios have been
proposed: the double-degenerate (DD, [5]) and the single
degenerate (SD, [6,7]) channel. The DD model involves the merger
of two CO WDs having a total mass larger than the Chandrasekhar
(Ch) mass limit [8,9]. In the SD model, the maximum stable mass of
the CO WD is $\sim 1.378 M_{\odot}$ (close to the Ch mass limit
[7]), and the companion is either a main sequence (or a slightly
evolved) star (WD+MS, [10-14]), or a red-giant star (WD+RG, [15])
or a He star (WD + He star, [16]). This scenario is supported by
many observations  [17] and has been widely studied by many groups
[18-24].

A direct way to confirm the progenitor model is to search for the
companion stars of SNe Ia in their remnants since they can survive
in the SD model but the entire system is destroyed after supernova
explosion in the DD system. The discovery of the potential
companion (Tycho G named by Ruiz-Lapuente et al. [25]) to Tycho's
supernova has verified the power of the method, and also the
reliability of the SD model. Recently, the chemical abundances
analysis of Tycho G by Gonz\'{a}lez-Hern\'{a}ndez [26] supproted
the findings of Ruiz-Lapuente et al. However, the observed
rotational velocity of Tycho G is not consistent with predictions
from an analytic theory. Predictions suggest that Tycho's
supernova companion should have a high rotation rate but the
observed rotational velocity is as low as 7.5 km s$^{\rm -1}$
[27]. It is therefore necessary to calculate the properties of the
companion stars from the SD model, which may be helpful in finding
the companions in supernova remnant, and compare them with the
observed properties of Tycho G to see if the predictions match
observations. Actually, some works has been done in this area
[18,20,21,28], but has been based on the WD+MS channel. In this
paper, the properties of the companion predicted in the WD+RG
channel are considered since this channel is dominant in early
type galaxies. As a result, our study should contribute to a more
complete picture of the SN Ia.

The paper is a following-up to our previous paper[29] and our
binary evolution calculation methods are similar to before(please
see [29] for more details, see also [30]). Therefore, we briefly
describe these method and then present our results in Section 2.
We end the paper with a discussion of the results in Section 3.

\section{Results}\label{sect:2}
In this paper, we calculate a series of binary evolutions with
different initial conditions for the WD mass, secondary mass and
orbital period. Both the WD + MS and WD + RG channels are
included. In an initial binary system, the companion fills its
Roche lobe at the MS or during the HG or RG stage, and then the
mass transfer occurs. The WD accretes the transferred material
from the companion, increasing its mass. Here, we adopt the
optically thick wind model suggested by Hachisu et al. [45]. We
simultaneously consider the mass-stripping effect by the wind and
the effect of a thermally unstable disk [18,38]. When the mass of
the WDs reaches $M_{\rm WD}^{\rm SN}=1.378 M_{\odot}$ (close to Ch
mass limit, [7]), we assume a SN Ia explosion occurs. We
subsequently record the state of the companion at the moment of
explosion. Now, based on observations, we may deduce various
parameters of the potential companion of a SN Ia, such as mass,
radius, effective temperature, luminosity, space velocity,
rotation velocity, etc. All these parameters may also be obtained
from our calculations.

It is important to note, however, that the companion's parameters
obtained at the moment of supernova explosion from our
calculations may be different from those found later in a SN Ia
remnant because the interaction with the supernova ejecta can
change the situation of the companion. During the interaction, the
ejecta may strip off some of the hydrogen-rich material from the
envelope of the companion. At the same time, the companion gains a
kick velocity, which is vertical to the orbital velocity. Part of
the kinetic energy of the ejecta is deposited into the envelope of
the companion. As a result, the radius and luminosity of the
companion will rise dramatically, and the companion may lose
dynamic equilibrium. After the interaction, the companion
reestablishes dynamic equilibrium quickly while it is still in the
process of moving back into thermal equilibrium. This process into
thermal equilibrium may last for $10^{\rm 3}$ - $10^{\rm 4}$ yr
[31-34]. The age of Tycho's supernova is about 440 yr, which means
that the suggested companion star, Tycho G, is still not back into
thermal equilibrium if it is the companion star of Tycho's
supernova. Similarly, a companion star is also not in thermal
equilibrium at the moment of a supernova explosion since mass
transfer is taking place before the explosion. So, considering
that the amount of stripped material from the companion is small
[35-37], we may assume that the parameters of a companion (such as
radius, effective temperature, mass and luminosity) are similar
before and after the SN Ia event. The space velocity of the
companion may also be assumed to be the same as its orbital
velocity since the kick velocity is much smaller than its orbital
velocity and as a result may be ignored [31,32,37]. Therefore, our
results can be compared directly with observations.

   \begin{figure}
   \centering
   \includegraphics[width=60mm,height=80mm,angle=270.0]{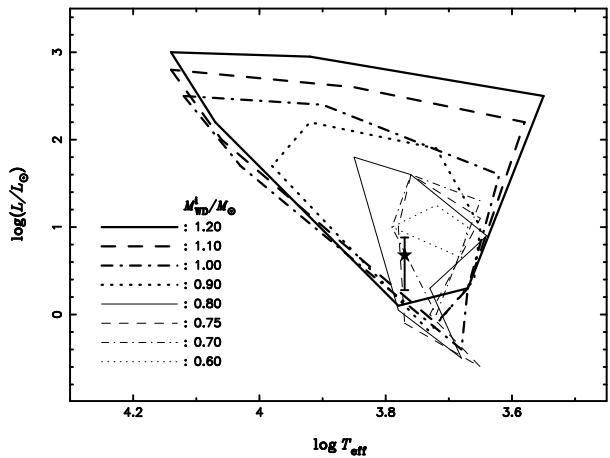}
   \caption{The parameter
spaces of luminosity and temperature of the companion at the
moment of supernova explosion with different initial WD masses.
The star represents the potential companion star of Tycho's
supernova, Tycho G, and the bar represents the observational error
[26].} %
    \end{figure}

\subsection{Luminosity and Effective Temperature of Companion at Explosion}\label{sect:2.1}
In Figure 1, we present the parameter spaces of luminosity and
temperature of the companion at the moment of supernova explosion
with different initial WD masses. The range in luminosity is very
large and the luminosity may reaches a maximum near $10^{\rm 3}
L_{\odot}$ since the initial mass of companion is as large as 5
$M_{\odot}$ and the WD+RG channel is also included. In addition,
the temperature range is much larger than that shown in Han
(2008)[28] since in this paper we considered the mass-stripping
effect by the optically thick wind and the effect of a thermally
unstable disk [18,38]. However, because an initial WD mass of
$\sim0.8M_{\odot}$ is favored based on detailed binary population
synthesis (BPS) results [14,29], most of the companions should be
located in the range of $3.7<\log T_{\rm eff}<3.8$ and $0<\log
L<2$, which is consistent with the BPS results of Han (2008)[28].
Since Han (2008)[28] only investigated the WD+MS channel, the
consistent result here implies that the contribution of the WD+RG
channel is not as important as the WD+MS channel [13,29], as shown
in [29]. The potential companion of Tycho's supernova, Tycho G, is
well within the range of highest probability in our range of
luminosity and temperature.

   \begin{figure}
   \centering
   \includegraphics[width=60mm,height=80mm,angle=270.0]{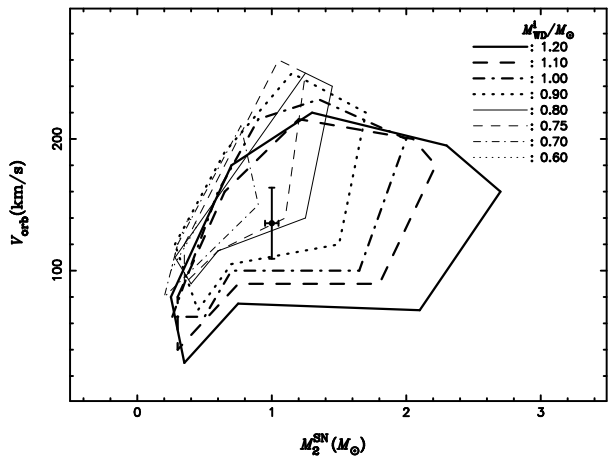}
   \caption{Similar to
Figure 1, but for companion mass and orbital velocity. The cross
represents Tycho G and the length of the cross shows the typical
observational errors, where the velocity of Tycho G here is space
velocity [25].} %
    \end{figure}

\subsection{Mass and Space Velocity}\label{sect:2.2}
Figure 2 shows the parameter spaces of companion mass and orbital
velocity for different initial WD masses at the moment of
explosion. The ranges of companion mass and orbital velocity here
are only slightly larger than those shown in Han (2008)[28] and
Meng \& Yang (2010) [21] even though we considered the
mass-stripping effect by optically thick wind and the effect of a
thermally unstable disk which can increase the maximum initial
companion mass from 3.5 $M_{\rm \odot}$ to as large as 5 $M_{\rm
\odot}$. However, the mass-stripping effect can strip off a large
amount of hydrogen-rich material from the envelope of the
companion, which may lead to a less massive companion than without
the effect at the moment of a SN Ia explosion. For the same reason
as in section 2.1, most of the companion stars should be in the
range of $0.5M_{\odot}<M_{\rm 2}^{\rm SN}<1.4M_{\odot}$ and
$100{\rm kms^{\rm -1}}<V_{\rm orb}<200{\rm kms^{\rm -1}}$, which
is also consistent with the BPS results of Han (2008)[28] and Meng
\& Yang (2010) [21]. It should again be noted that Han (2008)[28]
and Meng \& Yang (2010) [21] only investigated the WD+MS channel,
while in this study we also include the WD+RG channel. The
consistent results here mean that the contribution of the WD+RG
channel to SNe Ia is small and the number of companions with high
initial mass is smaller than that with low initial mass, as shown
in [29]. However, it should be emphasized that the WD+RG channel
is the main one contributing to very old SNe Ia in early type
galaxies and the progenitor systems with massive companion may be
the contributor to the young SNe Ia with age younger than 1Gyr.
Tycho G is also located in the parameter space with highest
probability. Then, these two parameters of Tycho G are also
consistent with the prediction from our model.

   \begin{figure}
   \centering
   \includegraphics[width=60mm,height=80mm,angle=270.0]{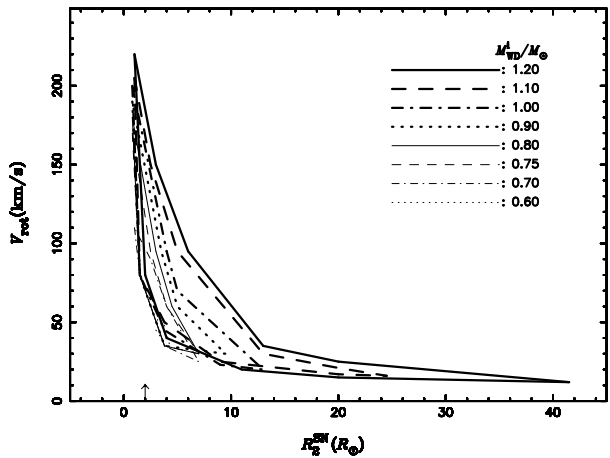}
   \caption{Similar to
Figure 1, but for companion radius and equatorial rotational
velocity. The arrow represents Tycho G [25,27].} %
    \end{figure}

\subsection{Companion Radius and Rotational Velocity}\label{sect:2.3}
We calculate the companion radius at the moment of explosion based
on the assumption that it is equal to its Roche lobe radius. The
equatorial rotation velocity is chosen as the rotational velocity
and we assume that the companion star co-rotates with its orbit.
In Figure 3, we show the parameter spaces of companion radius and
equatorial rotational velocity for different initial WD masses.
From the figure, we can see that the rotational velocity of the
surviving companions may be as high as 200 km s$^{\rm -1}$, which
means that their spectral lines should be noticeably broadened. In
general, a companion star with a large radius has a lower
rotational velocity and the rotation rate increases with
decreasing radius. The companion stars with radii larger than 10
$R_{\rm \odot}$ are all from the WD+RG channel and have low
rotational velocities. We expect companions in early type galaxies
to have these properties. The companion stars with radii smaller
than 10 $R_{\rm \odot}$ are mainly from the WD+MS channel. For the
reason mentioned in section 2.1, most of the companion stars
should should be in the vicinity $\sim100$ km s$^{\rm -1}$ and
$\sim1R_{\odot}$, which is also consistent with the results of Han
(2008)[28] and Meng \& Yang (2010) [21]. However, it should be
noted that although the radius of Tycho G matches well with our
calculation, its rotational velocity is much smaller than that
predicted from our model. Actually, this is the reason why
Kerzendorf et al. [27]\footnote{Note that the rotational velocity
of Tycho G was obtained based on an assumption that $\sin i$
(where $i$ is the inclination angle of rotational axis) is closer
to 1 than 0. } argued the companion nature of Tycho G for Tycho's
supernova (see discussion about this problem).

   \begin{figure}
   \centering
   \includegraphics[width=60mm,height=80mm,angle=270.0]{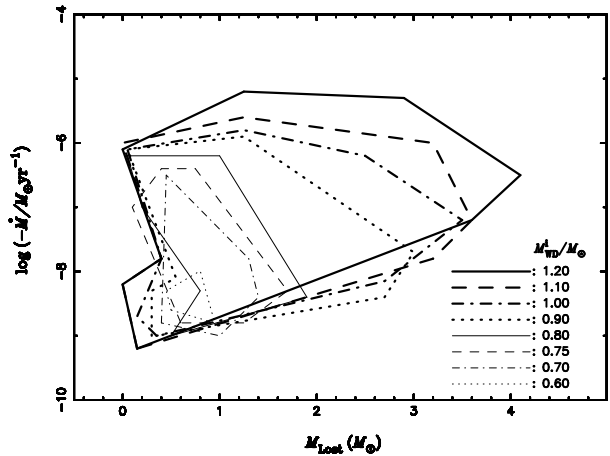}
   \caption{Similar to
Figure 1, but for mass-transfer rate at the moment of supernova
explosion and the total amount of material lost from the binary
system.} %
    \end{figure}

\subsection{Mass Loss}\label{sect:2.4}
Figure 4 shows the parameter spaces of mass-transfer rate at the
moment of supernova explosion and the total amount of material
lost from the binary system for different initial WD masses. From
the figure, one can see that some binary systems have lost a large
amount of material, while their mass-transfer rates are still very
high at the moment of supernova explosion, i.e. several times
$10^{\rm -6}M_{\odot}{\rm yr^{\rm -1}}$. The SNe Ia originating
from these stars may exhibit properties similar to those of SN
2002ic and 2006X [39,40], which may be experiencing an optically
thick wind phase before explosion [41]. However, supernovae
exploding during this phase are very rare [43]. In addition, there
are some systems in which the mass-transfer rates are very low
(lower than $10^{\rm -9}M_{\odot}{\rm yr^{\rm -1}}$) and almost no
material is lost. The lost material consists of two parts, i.e.
the optically thick wind from the surface of the WD and that which
is stripped off by the wind from the envelope of the companion.
The velocity of the wind is as high as 2000 km s$^{\rm -1}$, while
the velocity of those stripped off from the companion envelope is
only about 100 km s$^{\rm -1}$[18]. The lost material may be the
main origin of the color excess of SNe Ia [42]. For most cases,
the amount of material lost by the systems is around 1
$M_{\odot}$, and mass-transfer rate is around $10^{\rm
-8}M_{\odot}{\rm yr^{\rm -1}}\sim10^{\rm -7}M_{\odot}{\rm yr^{\rm
-1}}$. The results shown here may be helpful in finding progenitor
systems before a supernova explosion, in verify the potential
progenitor system from archive data after the supernova explosion,
or in constraining the X-ray luminosity of a progenitor system
candidate. The mass transfer rates can be converted to an X-ray
luminosity for the binary system via $L_{\rm
X}\sim\varepsilon|\dot{M}|$, where $\varepsilon=7\times10^{\rm
18}$ erg g$^{\rm -1}$ is the approximate amount of energy obtained
per gram of hydrogen converted into helium. The luminosity of the
X-ray source close to the site of SN 2007on was estimated to be
$(3.3\pm1.5)\times10^{\rm 37}{\rm erg} {\rm s}^{\rm -1}$ [44].
This luminosity corresponds to a mass accretion rate of
$\sim10^{\rm -7}M_{\odot}{\rm yr^{\rm -1}}$, which is consistent
with our calculations.

\section{Discussion and Conclusion}\label{sect:3}
A good way of discriminating between the many SN Ia progenitor
scenarios is to search for the companion of a SN Ia in its
remnant. Unless the companion is another WD (DD channel, in which
it has been destroyed by the mass-transfer process itself before
explosion), it survives and shows some special properties in its
spectrum, which originates from the contamination of the supernova
ejecta [25,31]. Tycho's supernova, which is one of only two SNe Ia
observed in our Galaxy, provides an opportunity to observationally
address the identification of the surviving companion. By
searching the region of its remnant, a sun-like star, Tycho G, was
suggested to be the companion of Tycho's supernova. Chemical
abundance analysis of Tycho G upholds its companion nature[26].
Interestingly, some integral properties of Tycho G (the mass,
space velocity, radius, luminosity and effective temperature) are
all consistent with our computational results, with the exception
of the rotational velocity (see also [27]). However, whether the
inconsistent result for the rotational velocity is a key factor
against for the companion nature of Tycho G still should be
investigated more carefully, because the interaction between
supernova ejecta and companion star is still unclear. For example,
some calculations predicted that there should be a large amount of
hydrogen-rich material stripped from the companion envelope by the
ejecta in the supernova remnant. This stripped materials may be
revealed by narrow H$_{\rm \alpha}$ lines in later-time spectra of
SNe Ia [31-33], but this prediction has not been uphold by
observations [36,37]. Whether the interaction may change the
rotational velocity of the companion star is unclear since the
rotational property of the companion has never been considered
when simulating the interaction between the supernova ejecta and
the companion. For example, because of rotation, the amount of
supernova ejecta accreted by the companion on the blue-shift side
(rotational velocity is facing the ejecta velocity) could be
slightly larger than that on the red-shift side. The small
discrimination in angular momentum between the blue- and red-shift
side of the companion caused by the different accretion rates may
change the rotational dynamics of the companion. If the
interaction slows the rotational velocity of the companion,Tycho G
is fully explained by the SD model making it an excellent
candidate for the companion Tucho's supernova. Otherwise, the
companion nature of Tycho G must be reconsidered. So, more
detailed study of Tycho G is encouraged. Furthermore, when one
simulates the interaction between supernova ejecta and companion,
the rotational property should be considered carefully.

\baselineskip 18pt
 {\textbf{Acknowledgement:} \emph{This work was
partly supported by the Natural Science Foundation of China (Grant
No. 11003003), the Project of Science and Technology from the
Ministry of Education (211102) and the China Postdoctoral Science
Foundation funded project 20100480222. }}

\textbf{Reference}

[1]Riess A, Filippenko A V, Challis P, et al. Observational
evidence from supernovae for an accelerating universe and a
Cosmological Constant. Astron J, 1998, 116: 1009--1038

[2]Perlmutter S, Aldering G, Goldhaber G, et al. Measurements of
Omega and Lambda from 42 high-redshift supernovae. Astrophys J,
1999, 517: 565--586

[3]Hillebrandt W, Niemeyer J C. Type Ia supernova explosion
models. Ann Rev Astron Astrophys, 2000, 38: 191--230

[4]Leibundgut B. Type Ia Supernovae. Astron Astrophys Rev, 2000,
10: 179--20

[5]Iben I, Tutukov A V. Supernovae of type I as end products of
the evolution of binaries with components of moderate initial mass
(M not greater than about 9 solar masses). Astrophys J Sup, 1984,
54: 335--372

[6]Whelan J, Iben I. Binaries and Supernovae of Type I. Astrophys
J, 1973, 186: 1007--1014

[7]Nomoto K, Thielemann F-K, Yokoi K. Accreting white dwarf models
of Type I supernovae III - Carbon deflagration supernovae.
Astrophys J, 1984, 286: 644--658

[8]Webbink R F. Double white dwarfs as progenitors of R Coronae
Borealis stars and Type I supernovae. Astrophys J, 1984, 277:
355--360

[9]Han Z. The formation of double degenerates and related objects.
Mon Not Roy Astron Soc, 1998, 296: 1019--1040

[10]Li X D,  van den Heuvel E P J. Evolution of white dwarf
binaries: Supersoft X-ray sources and progenitors of Type Ia
supernovae. Astron Astrophys, 1997, 322: L9--L12

[11]Han Z, Podsiadlowski Ph. The single-degenerate channel for the
progenitors of Type Ia supernovae. Mon Not Roy Astron Soc, 2004,
350: 1301--1309

[12]Chen W C, Li X D. On the progenitors of super-Chandrasekhar
mass Type Ia supernovae. Astrophys J, 2009, 702: 686--691

[13]Wang B, Li X D, Han Z. The progenitors of Type Ia supernovae
with long delay times. Mon Not Roy Astron Soc, 2010, 401:
2729--2738

[14]Meng X, Chen X, Han Z. A single-degenerate channel for the
progenitors of Type Ia supernovae with different metallicities.
Mon Not Roy Astron Soc, 2009, 395: 2103--2116

[15]L\"{u} G, Zhu C, Wang Z, et al. An alternative symbiotic
channel to Type Ia supernovae. Mon Not Roy Astron Soc, 2009, 396:
1086--1095

[16]Wang B, Meng X, Chen X, et al. The helium star donor channel
for the progenitors of Type Ia supernovae. Mon Not Roy Astron Soc,
2009, 395: 847--854

[17]Parthasarathy M, Branch D, Jeffery D J, Baron E. Progenitors
of type Ia supernovae: Binary stars with white dwarf companions.
New Astron Rev, 2007, 51: 524--538

[18]Hachisu I, Kato M, Nomoto K. Young and massive binary
progenitors of Type Ia supernovae and their circumstellar matter.
Astrophys J, 2008, 679: 1390--1404

[19]Langer N, Deutschmann A, Wellstein S, et al. The evolution of
main sequence star + white dwarf binary systems towards Type Ia
supernovae. Astron Astrophys, 2000, 362: 1046--1064

[20]Meng X, Yang W, Geng X. WD+MS Systems as Progenitors of Type
Ia Supernovae with Different Metallicities. Publ Astron Soc Jpn,
2009, 61: 1251--1260

[21]Meng X, Yang W. Companion stars of Type Ia supernovae with
different metallicities. Mon Not Roy Astron Soc, 2010, 401:
1118--1130

[22]Meng X, Yang W. The envelope mass of red giant donors in Type
Ia supernova progenitors. Astron Astrophys, 2010, 516: A47--A51

[23]Chen W C, Li X D. Evolving to Type Ia Supernovae with Long
Delay Times. Astrophys J, 2007, 658: L51--L54

[24]Wang B, Liu Z, Han Y, et al. Birthrates and delay times of
Type Ia supernovae. Sci Chi G, 2010, 53: 586--590

[25]Ruiz-Lapuente P, et al. The binary progenitor of Tycho Brahe's
1572 supernova. Nature, 2004, 431: 1069--1072

[26]Gonz\'{a}lez-Hern\'{a}ndez J, Ruiz-lapuente P, Filippenko A,
et al. The Chemical Abundances of Tycho G in Supernova Remnant
1572. Astrophys J, 2009, 691: 1--15

[27]Kerzendorf W E, Schmidt B P, Asplund M, et al. Subaru
High-Resolution Spectroscopy of Star G in the Tycho Supernova
Remnant. Astrophys J, 2009, 701: 1665--1672

[28]Han Z. Companion Stars of Type Ia Supernovae. Astrophys J,
2008, 677: L109--L112

[29]Meng X, Yang W. A Comprehensive Progenitor Model for SNe Ia.
Astrophys J, 2010, 710: 1310--1323

[30]Meng X, Yang W, Li Z. The initial and final state of SNe Ia
from the single degenerate model. Sci Chi G, 2010, 53, 1732--1738

[31]Marietta E, Burrows A, Fryxell B. Type IA Supernova Explosions
in Binary Systems: The Impact on the Secondary Star and Its
Consequences. Astrophys J Suppl S, 2000, 128: 615--650

[32]Meng X, Chen X, Han Z. The Impact of Type Ia Supernova
Explosions on the Companions in a Binary System. Publ Astron Soc
Jpn, 2007, 59: 835--840

[33]Pakmor R, R\"{o}opke F K, Weiss A, Hillebrandt W. The impact
of type Ia supernovae on main sequence binary companions. Astron
Astrophys, 2008, 489: 943--951

[34]Meng X, Yang W. Companion stars of Type Ia supernovae with
different metallicities. Mon Not Roy Astron Soc, 2010, 401:
1118--1130

[35]Mattila S, Lundqvist P, Sollerman J, et al. Early and late
time VLT spectroscopy of SN 2001el - progenitor constraints for a
type Ia supernova. Astron Astrophys, 2005, 443: 649--662

[36]Leonard D C. Constraining the Type Ia Supernova Progenitor:
The Search for Hydrogen in Nebular Spectra. Astrophys J, 2007,
670: 1275--1282

[37]Lu F J, et al. The Single-degenerate Binary Origin of Tycho's
Supernova as Traced by the Stripped Envelope of the Companion.
arXiv: 1102.3829

[38]Xu X, Li X. Evolution of long-period, white-dwarf binaries:
application to GRO J1744-28 and type Ia supernovae. Astron
Astrophys, 2009, 495: 243--248

[39]Hamuy M, et al. An asymptotic-giant-branch star in the
progenitor system of a type Ia supernova. Nature, 2003, 424:
651--654

[40]Patat F, Chandra P, Chevalier R, et al. Detection of
circumstellar material in a normal Type Ia supernova. Science,
2007, 317: 924--926

[41] Han Z, Podsiadlowski Ph. A single-degenerate model for the
progenitor of the Type Ia supernova 2002ic. Mon Not Roy Astron
Soc, 2006, 368: 1095--1100

[42]Meng X, Chen X, Han Z, Yang W. Color excesses of type Ia
supernovae from the single-degenerate channel model. Res Astron
Astrophys, 2009, 9: 1259--1269

[43]Meng X, Yang W, Geng X. Is SN 2006X from a WD + MS system with
optically thick wind?. New Astron, 2010, 15:343--345

[44]Voss R, Nelemans G. Discovery of the progenitor of the type Ia
supernova 2007on. Nature, 2008, 451: 802--804

[45]Hachisu I, Kato M, Nomoto K. A New Model for progenitor
Systems of Type Ia Supernovae. Astrophys J, 1996, 470: L97--L100
\end{document}